\documentclass{article}
\usepackage{graphics}
\usepackage{epsfig}

\begin{document}

\begin{center} {\large \bf Collision-induced stimulated photon echo at the
transition 0-1}
\end{center}
\bigskip
\begin{center}
\textbf{V. A. Reshetov, E. N. Popov}\\
\bigskip
\textit{Department of General and Theoretical Physics, Tolyatti
State University, 14 Belorousskaya Street, 446667 Tolyatti, Russia}
\end{center}

\begin{abstract}
The stimulated photon echo formed by the sequence of three laser
excitation pulses in gaseous medium on the transition with the
angular momentum change $J_{a}=0 \rightarrow J_{b}=1$ is considered
with an account of elastic depolarizing collisions. The polarization
properties of this echo may be potentially interesting for the
purposes of data processing. It is shown, that with the polarization
of the first pulse orthogonal to that of the second and third pulses
the appearance of the echo signal will be entirely determined by the
action of collisions. The existence of this echo is determined by
the difference in the alignment $\Gamma_{b}^{(2)}$ and the
orientation $\Gamma_{b}^{(1)}$ relaxation rates of the excited
level.
\end{abstract}

\section{Introduction}

Since its first observation \cite{n1} photon echo was widely
employed for the study of fast relaxation processes in various media
(see, for example, some recent papers \cite{n2,n3,n4,n5}). It is
also considered as a strong contender for implementation of quantum
memory (see some recent reviews \cite{n6,n7} and the papers
\cite{n8,n9,n10,n11,n12} in the special issue of Journal of Physics
B, dedicated to quantum memories). Photon echo proved to be a
versatile tool due to the large number of its modifications suitable
for various purposes. The stimulated photon echo is formed on an
optically-allowed transition $a\rightarrow b$ by the sequence of
three resonant laser pulses. With the lower resonant level being
some metastable level, the time interval between the excitation
pulses may be rather long making such long-lived stimulated photon
echoes attractive for purposes of data storage \cite{n13,n14,n15}.
The modified stimulated photon echo \cite{n16,n17} is formed when
the first two excitation pulses are in resonance with some
optically-allowed transition $a\rightarrow b$, while the third pulse
is in resonance with some adjacent transition $b\rightarrow c$. This
echo signal is perfectly suited for the measurement of relaxation
characteristics of the intermediate level $b$ \cite{n18}. Another
type of echo, which may be generated in a three-level system by
three excitation pulses is the tri-level echo \cite{n19}. This echo
is formed, when the first and the third excitation pulses are in
resonance with one of the two optically-allowed transitions
$a\rightarrow b$ sharing common level $a$, while the second pulse is
in resonance with the other transition $a\rightarrow c$. Such echoes
provide information about the relaxation characteristics of the
optically-forbidden transition $b\rightarrow c$ \cite{n20}. A
variety of novel photon echo schemes were elaborated for the
implementation of quantum memories
\cite{n21,n22,n23,n24,n25,n26,n27,n28,n29}.

The collisional relaxation of photon echoes in gases was intensively
studied both theoretically and experimentally starting from their
first observation in \cite{n30} (see, for example,
\cite{n31,n32,n33,n34}). Generally, the relaxation processes act to
decrease the amplitude of the echo signals, however in some cases,
when the conventional echo is locked, the relaxation may act to
break the lock and to produce the relaxation-induced echo signals.
Such collision-induced two-pulse photon echo was recently observed
in the ytterbium vapour on the transition $J_{a}=0 \rightarrow
J_{b}=1$ \cite{n35}. This echo appears due to the difference in the
relaxation rates of the two components of the atomic dipole moment
-- collinear with the atomic velocity and perpendicular to it. The
possibility of observation of the collision-induced stimulated
photon echoes was indicated in papers \cite{n36,n37}. This type of
echoes appear due to the difference in the relaxation rates of
various multipole momenta of resonant levels. In the experiment
\cite{n17} the intensity of the modified stimulated photon echo
formed in ytterbium vapour on the transitions $J_{a}=0 \rightarrow
J_{b}=1 \rightarrow J_{c}=1$ was reduced greatly when the
polarizations of all the three excitation pulses became collinear,
which was interpreted in \cite{n38} as a collision-induced echo.
Inspired by the experiment \cite{n35} we propose here a scheme, much
simpler than in \cite{n36,n37}, for observation of the
collision-induced stimulated photon echo on a single transition
$J_{a}=0 \rightarrow J_{b}=1$. In this case only three
non-degenerate states are involved in the echo formation - the lower
level and the two substates of the upper level, so that the
stimulated photon echo generally contains all the types of echoes
characteristic for three-level systems -- the conventional two-level
stimulated echo, the modified stimulated echo and the tri-level
echo. Due to the simplicity of the level structure the choice of
polarizations of the three excitation pulses provides enough freedom
to separate any of these three echo types or to lock all of them
enabling the appearance of the collision-induced echoes. Such
polarization properties make this echo scheme potentially
interesting for the purposes of data processing, for implementation
of single q-bit quantum gates, in particular. However this echo
scheme is not valid for data storage, since the ground state is
non-degenerate ($J_{a}=0$) and the storage time is limited by the
rather short radiative lifetime of the excited level \cite{n39}. The
appearance of the collision-induced echoes in this case is entirely
determined by the difference in the alignment and orientation
relaxation rates of the excited level, so its observation will
provide information about depolarizing collisions.

In sections 2 and 3 the atomic dynamics under the action of short
laser pulses and under the action of elastic depolarizing collisions
and spontaneous radiational decay is described. In section 4 the
general expression for the electric field strength of the echo
signal is derived. In section 5 the polarization properties of the
echo signal are analyzed in the absence of collisions and the
possibilities of data processing is discussed, while in section 6
the polarization properties of the collision-induced stimulated
photon echo are studied.

\section{Atomic dynamics under the action of laser pulses}

Let us consider the stimulated photon echo formation on the
transition with the angular momentum change $J_{a}=0 \rightarrow
J_{b}=1$ by the three resonant laser pulses with the durations
$T_{1}$, $T_{2}$ and $T_{3}$, $\tau_{1}$ being the time interval
between the first and the second pulses  and $\tau_{2}$ -- between
the second and the third. The electric field strength of these
pulses, propagating along the axis $Z$ with the carrier frequency
$\omega$, may be put down as:
     \begin{equation}\label{q1}
  \textbf{E}_{n}=e_{n}\textbf{l}_{n}\exp\{-i(\omega
  t-kz)\}+c.c.,~n=1,2,3,
       \end{equation}
where $e_{n}$ and $\textbf{l}_{n}$ are the amplitudes and unit
polarization vectors of the pulses. We shall consider the pulses to
be much shorter, than all the relaxation times. Then the dynamics of
the atom under the action of n-th laser pulse is governed by the
equation for the slowly-varying atomic density matrix $\hat{\rho}$:
      \begin{equation}\label{q2}
  \dot{\hat{\rho}}=i\left[\hat{\Omega}_{n},\hat{\rho}\right],~
\hat{\Omega}_{n}=
\frac{|d|e_{n}}{\hbar}\left(\hat{g}_{n}+\hat{g}_{n}^{\dag}\right).
      \end{equation}
Here $d=d(J_{a}J_{b})$ is the reduced matrix element of the electric
dipole moment operator for the transition $J_{a}=0\rightarrow
J_{b}=1$, while $\hat{g}_{n}=(\hat{\textbf{g}}\textbf{l}_{n}^{*})$,
where $\hat{\textbf{g}}$ is the dimensionless electric dipole moment
operator for this transition, the matrix elements of its circular
components, expressed through Wigner 3J-symbols, are as follows:
      \begin{equation}\label{q3}
(\hat{g}_{n})^{ab}_{0q}=\frac{1}{\sqrt{3}}(l_{n,-q})^{*},
    \end{equation}
where $l_{n,q}$ are the circular components of the unit polarization
vector of the n-th laser pulse. The solution of the equation
(\ref{q2}) may be expressed through the evolution operator
$\hat{S}_{n}$:
    \begin{equation}\label{q4}
\hat{\rho}(T_{n})=\hat{S}_{n}\hat{\rho}(0)\hat{S}^{\dag}_{n},~
\hat{S}_{n}=\exp\left(i\int_{0}^{T_{n}}\hat{\Omega}_{n}dt\right),
    \end{equation}
$\hat{\rho}(0)$ and $\hat{\rho}(T_{n})$ being the atomic density
matrices at some point $z$ of the gaseous medium before and after
the n-th laser pulse passes through this point. The explicit
expression of the evolution operator may be obtained by means of
expansion of the exponent function in Taylor series and by
diagonalization of the operator $\hat{\Omega}_{n}$. Defining the
area of the n-th excitation pulse by the relation
     \begin{equation}\label{q5}
\theta_{n}=\frac{2|d|}{\sqrt{3}\hbar}\int_{0}^{T_{n}}e_{n}dt,
     \end{equation}
we immediately obtain the matrix elements of the evolution operator
involving the lower atomic level $a$:
    \begin{equation}\label{q6}
(\hat{S}_{n})^{aa}_{00} = \cos\left(\frac{\theta_{n}}{2}\right),
(\hat{S}_{n})^{ab}_{0q} =
i\sin\left(\frac{\theta_{n}}{2}\right)(l_{n,-q})^{*}.
     \end{equation}
In order to obtain the matrix elements $(\hat{S}_{n})^{bb}$ of the
evolution operator referring to the upper state $b$, we note that
the operator $\hat{g}_{n}^{\dag}\hat{g}_{n}$  has the following
three eigenvectors: the "bright" state
     \begin{equation}\label{q7}
|b_{n}>=\sum_{q=\pm 1}l_{n,-q}|J_{b}=1,m_{b}=q>
     \end{equation}
with non-zero eigenvalue, the "dark" state
     \begin{equation}\label{q8}
|\tilde{b}_{n}>=\sum_{q=\pm 1}s_{n,-q}|J_{b}=1,m_{b}=q>
     \end{equation}
with zero eigenvalue, where $\textbf{s}_{n}$ is the unit vector in
the plane $XY$, which is orthogonal to the vector $\textbf{l}_{n}$
($\textbf{s}_{n}\textbf{l}_{n}^{*}=0$), and the state
$|J_{b}=1,m_{b}=0>$, which is not involved into interactions at any
polarization of laser pulse and may be neglected. Then, for the
matrix elements of the part $(\hat{S}_{n})^{bb}$ of the evolution
operator we obtain:
     \begin{equation}\label{q9}
(\hat{S}_{n})^{bb}_{qq'} = s_{n,-q}s_{n,-q'}^{*} +
l_{n,-q}l_{n,-q'}^{*}\cos\left(\frac{\theta_{n}}{2}\right).
     \end{equation}

\section{Atomic dynamics under the action of depolarizing
collisions}

The evolution of the density matrix elements in the time interval
between the pulses is determined by the frequency detuning
$\Delta=kv_{z}-\omega+\omega_{0}$, including Doppler shift $kv_{z}$,
and by the irreversible relaxation. Here we shall take into account
the two most rapid relaxation processes - the spontaneous radiation
decay of the excited level and the elastic depolarizing collisions,
which do not change the atomic velocities but give rise to the
transitions between various Zeeman sublevels of atomic resonant
levels. Then, the equations for the irreducible density matrix
components ($\alpha,\beta=a,b$)
$$\psi^{\alpha\beta}_{kq} = \sqrt{2J_{\alpha}+1} \sum_{\mu,\nu}
(-1)^{J_{\alpha}-\mu} \left( \matrix{J_{\alpha}&J_{\beta}&k \cr \mu
&-\nu & q}\right)\rho^{\alpha\beta}_{\mu\nu}$$ are as follows
\cite{n34}:
     \begin{equation}\label{q10}
\dot{\psi}^{bb}_{kq} = - \gamma_{b}^{(k)}\psi^{bb}_{kq},
     \end{equation}
     \begin{equation}\label{q11}
\dot{\psi}^{aa}_{kq} = - \gamma_{a}^{(k)}\psi^{aa}_{kq} +
\Gamma_{ab}^{(k)}\psi^{bb}_{kq},
     \end{equation}
     \begin{equation}\label{q12}
\dot{\psi}^{ab}_{kq} = -
[\gamma^{(k)}+i(\Delta^{(k)}-\Delta)]\psi^{ab}_{kq},
     \end{equation}
where
$$\gamma_{a,b}^{(k)} = \gamma_{a,b}^{(0)} + \Gamma_{a,b}^{(k)},$$
$$\gamma^{(k)} = \frac{1}{2}
\left(\gamma_{a}^{(0)}+\gamma_{b}^{(0)}\right) + \Gamma^{(k)},$$
$$\Gamma_{ab}^{(k)} =(-1)^{k+1+J_{a}+J_{b}}\gamma_{ab}\times$$
$$\sqrt{(2J_{a}+1)(2J_{b}+1)} \left\{ \matrix{J_{a}&J_{a}&k \cr
J_{b}&J_{b}&1}\right\}.$$ Here $1/\gamma_{a}^{(0)}$ and
$1/\gamma_{b}^{(0)}$ are the lifetimes of atomic levels $a$ and $b$
due to spontaneous radiation decay, $\Gamma_{a}^{(k)}$ and
$\Gamma_{b}^{(k)}$ describe the relaxation of the levels $a$ and $b$
due to the action of the elastic depolarizing collisions,
($\Gamma_{a}^{(0)}=\Gamma_{b}^{(0)}=0$, since the elastic collisions
do not alter the population of the levels), $\Gamma^{(k)}$ and
$\Delta^{(k)}$ describe the relaxation of the irreducible components
of the optical coherence matrix due to elastic depolarizing
collisions, and $1/\gamma_{ab}$ is the partial lifetime of the upper
level $b$ due to the spontaneous radiation transitions to the lower
level $a$, while Wigner 6J-symbols are denoted in a usual way
\cite{n40}. In case of transition $J_{a}=0 \rightarrow J_{b}=1$ in
ytterbium vapour $a$ is the ground state, so $\gamma_{a}^{(0)}=0$,
$\gamma_{b}^{(0)}=\gamma_{ab}=\gamma$. The solution of the equations
(\ref{q10})-(\ref{q12}) is easily obtained and the elements
$\rho^{\alpha\beta}_{\mu\nu}(t)$ of the density matrix in the
original basis of Zeeman states, expressed through these solution by
means of inverse transformation of irreducible density matrix
components are as follows:
     \begin{equation}\label{q13}
\rho^{aa}_{00}(t) = e^{-\gamma t}\rho^{aa}_{00}(0),
     \end{equation}
     \begin{equation}\label{q14}
\rho^{ab}_{0q}(t) = e^{- (\gamma^{(1)}-i\delta)t} \rho^{ab}_{0q}(0),
     \end{equation}
     \begin{equation}\label{q15}
\rho^{bb}_{qq'}(t) = e^{-\gamma t}
\sum_{\sigma,\sigma'}D^{\sigma\sigma'}_{qq'}(t)
\rho^{bb}_{\sigma\sigma'}(0),
     \end{equation}
     \begin{equation}\label{q16}
D^{\sigma\sigma'}_{qq'}(t)=\delta_{q\sigma}\delta_{q'\sigma'} +
R^{\sigma\sigma'}_{qq'}(t).
     \end{equation}
where $\delta=\Delta - \Delta^{(1)}$,
$$R^{1,1}_{1,1}(t) = R^{-1,-1}_{-1,-1}(t) = \frac{1}{6}h_{b}^{(2)}(t) +
\frac{1}{2}h_{b}^{(1)}(t),$$
$$R^{1,1}_{-1,-1}(t) = R^{-1,-1}_{1,1}(t) = \frac{1}{6}h_{b}^{(2)}(t) -
\frac{1}{2}h_{b}^{(1)}(t),$$
$$R^{1,-1}_{1,-1}(t) = R^{-1,1}_{-1,1}(t) = h_{b}^{(2)}(t),$$
     \begin{equation}\label{q17}
h_{b}^{(k)}(t)= 1 - e^{-\Gamma_{b}^{(k)}t},
     \end{equation}
while all the other elements of relaxation matrix
$R^{\sigma\sigma'}_{qq'}(t)$ are zero.

\section{Stimulated photon echo formation}

Initially the atom is at its ground state $a$. The first laser pulse
creates atomic coherence on the transitions $m_{a}=0\rightarrow
m_{b}=q=\pm 1$, which is described by the non-diagonal elements of
the atomic density matrix according to (\ref{q4}):
     \begin{equation}\label{q18}
\rho^{ab}_{0q}(T_{1})=(\hat{S}_{1})^{aa}_{00}
(\hat{S}^{\dag}_{1})^{ab}_{0q}.
     \end{equation}
The evolution of these non-diagonal density matrix elements, which
are responsible for the echo formation, in the time interval between
the pulses is determined by the equation (\ref{q14}), so that at the
instant of time $t-z/c-T_{1}=\tau_{1}$, when the second excitation
pulse arrives at point $z$ of the gaseous medium,
     \begin{equation}\label{q19}
\rho^{ab}_{0q}(\tau_{1})=\rho^{ab}_{0q}(T_{1})
e^{-\gamma^{(1)}\tau_{1}+i\delta\tau_{1}}.
     \end{equation}
The second laser pulse creates the modulated velocity distributions,
proportional to $e^{-i\delta\tau_{1}} \sim e^{-ikv_{z}\tau_{1}}$, of
the atoms at the upper and lower levels according to the solution of
the equation (\ref{q4}):
    \begin{equation}\label{q20}
\rho^{aa}_{00}(T_{2})=\sum_{q}(\hat{S}_{2})^{ab}_{0q}
\rho^{ba}_{q0}(\tau_{1}) (\hat{S}^{\dag}_{2})^{aa}_{00},
    \end{equation}
    \begin{equation}\label{q21}
\rho^{bb}_{qq'}(T_{2})=\sum_{q''}(\hat{S}_{2})^{bb}_{qq''}
\rho^{ba}_{q''0}(\tau_{1}) (\hat{S}^{\dag}_{2})^{ab}_{0q'}.
    \end{equation}
After the passage of the second pulse through the medium the
relaxation of the these modulated velocity distributions is
described by the equations (\ref{q13}) and (\ref{q15}), so that at
the instant of time $t-z/c-T_{1}-\tau_{1}-T_{2}=\tau_{2}$, when the
third excitation pulse arrives at point $z$ of the medium:
     \begin{equation}\label{q22}
\rho^{aa}_{00}(\tau_{2}) = e^{-\gamma
\tau_{2}}\rho^{aa}_{00}(T_{2}),
     \end{equation}
     \begin{equation}\label{q23}
\rho^{bb}_{qq'}(\tau_{2}) = e^{-\gamma \tau_{2}}
\sum_{\sigma,\sigma'}D^{\sigma\sigma'}_{qq'}(\tau_{2})
\rho^{bb}_{\sigma\sigma'}(T_{2}).
     \end{equation}
The third laser pulse creates atomic coherence on the transitions
$m_{a}=0\rightarrow m_{b}=q=\pm 1$ from these modulated velocity
distributions (\ref{q22}) and (\ref{q23}) according to (\ref{q4}):
     \begin{equation}\label{q24}
\rho^{ab}_{0q}(T_{3}) = [\rho^{ab}_{0q}(T_{3})]_{a} +
[\rho^{ab}_{0q}(T_{3})]_{b},
     \end{equation}
$$ [\rho^{ab}_{0q}(T_{3})]_{a} =
(\hat{S}_{3})^{aa}_{00}\rho^{aa}_{00}(\tau_{2})
(\hat{S}^{\dag}_{3})^{ab}_{0q},$$
$$[\rho^{ab}_{0q}(T_{3})]_{b} =
\sum_{q',q''}(\hat{S}_{3})^{ab}_{0q'}\rho^{bb}_{q'q''}(\tau_{2})
(\hat{S}^{\dag}_{3})^{bb}_{q''q}.$$ The evolution of these density
matrix elements (\ref{q24}) after the action of the third laser
pulse is determined by the equation (\ref{q14}), so that at the
instant of time $t'=t-z/c-T_{1}-\tau_{1}-T_{2}-\tau_{2}-T_{3}$:
     \begin{equation}\label{q25}
\rho^{ab}_{0q}(t')=\rho^{ab}_{0q}(T_{3}) e^{-\gamma^{(1)}t'+i\delta
t'}.
     \end{equation}
The electric field strength of the stimulated photon echo signal
$$ \textbf{E}^{e}=\textbf{e}^{e}(t')\exp\{-i(\omega t-kz)\}+c.c.,$$
is obtained from the Maxwell equations in a usual way:
       \begin{equation}\label{q26}
\textbf{e}^{e}(t') =  i e_{0}\int f(\textbf{v})d\textbf{v}
Tr\{\hat{\rho}(t')\hat{\textbf{g}}\},
       \end{equation}
where $e_{0}= 2\pi\omega L n_{0}|d|/c$, $L$ is the length of the
gaseous medium, $n_{0}$ is the concentration of resonant atoms,
$f(\textbf{v})$ is the Maxwell velocity distribution function, while
the atomic density matrix $\hat{\rho}(t')$ at the instant of time
$t'$ is determined by (\ref{q25}). Henceforth we shall neglect the
action of irreversible relaxation during the echo pulse as well as
it was neglected during the excitation pulses and confine ourselves
to the case of exact resonance. After the successive substitution of
(\ref{q18})-(\ref{q25}) in (\ref{q26}) with an account of the
expressions (\ref{q6}) and (\ref{q9}) for the evolution operator the
projections $e^{e}_{n}(t')=\textbf{e}^{e}(t')\textbf{e}_{n}^{*}$ of
the echo amplitude on the two orthonormal vectors $\textbf{e}_{n}$
($n=1,2$) in the plane of polarization $XY$ may be expressed by the
following equation:
       \begin{equation}\label{q27}
e_{n}^{e}(t') =e_{0}\frac{1}{\sqrt{3}}
e^{-2\gamma^{(1)}\tau_{1}-\gamma\tau_{2}}I_{e}(t') G_{n}^{e},
       \end{equation}
where factor
       \begin{equation}\label{q28}
I_{e}(t')= e^{-\frac{1}{4}k^{2}u^{2}(t'-\tau_{1})^{2}}
       \end{equation}
describes the temporal shape of the echo pulse, $u$ being the atomic
thermal velocity, while factor
       \begin{equation}\label{q29}
G_{n}^{e}=\frac{1}{2}\sin(\theta_{1})(F_{n}^{a}+F_{n}^{b}+F_{n}^{c})
       \end{equation}
describes the polarization properties of the echo signal. This
factor contains three terms, the first two of them $F_{n}^{a}$ and
$F_{n}^{b}$  are responsible for the contributions of the coherence
created by the first two exciting pulses at the atomic levels $a$
and $b$ respectively:
       \begin{equation}\label{q30}
F_{n}^{a} = \frac{1}{4}\sin(\theta_{2}) \sin(\theta_{3})
(\textbf{l}_{1}^{*}\textbf{l}_{2})
(\textbf{e}_{n}^{*}\textbf{l}_{3}),
       \end{equation}
       \begin{equation}\label{q31}
F_{n}^{b} = \sin\left(\frac{\theta_{2}}{2}\right)
\sin\left(\frac{\theta_{3}}{2}\right)
U_{2}(\textbf{l}_{3},\textbf{l}_{1})
U_{3}^{*}(\textbf{e}_{n},\textbf{l}_{2}),
       \end{equation}
       \begin{equation}\label{q32}
U_{k}(\textbf{a},\textbf{b}) =
(\textbf{a}\textbf{s}_{k}^{*})(\textbf{s}_{k}\textbf{b}^{*}) +
\cos\left(\frac{\theta_{k}}{2}\right)
(\textbf{a}\textbf{l}_{k}^{*})(\textbf{l}_{k}\textbf{b}^{*}),
       \end{equation}
while the third term $F_{n}^{c}$ is entirely determined by the
action of elastic depolarizing collisions.

\section{Photon echo polarization in the absence of collisions}

In the absence of elastic depolarizing collisions $\Gamma_{b}^{(1)}$
= $\Gamma_{b}^{(2)}$ = 0, the relaxation matrix
$R_{qq'}^{\sigma\sigma'}(\tau_{2})$ in (\ref{q16}) becomes zero and
the echo signal is determined only by the terms $F_{n}^{a}$ and
$F_{n}^{b}$ in the expression (\ref{q29}). In the general expression
(\ref{q29}) three different types of echo signals may be
distinguished.

The first one occurs when all the excitation pulses are identically
polarized: $\textbf{l}_{1}$ = $\textbf{l}_{2}$ = $\textbf{l}_{3}$ =
$\textbf{e}_{1}$. All the three pulses couple the non-degenerate
ground state to a single substate $|b_{1}>$ ("bright state") of the
upper level (\ref{q7}). In this case the echo signal represents
itself a conventional two-level stimulated photon echo, its
polarization coincides with that of the excitation pulses:
       \begin{equation}\label{q33}
G_{1}^{e}=\frac{1}{4}\sin(\theta_{1})\sin(\theta_{2})\sin(\theta_{3}),~
G_{2}^{e}=0,
       \end{equation}
and the maximum echo amplitude takes place at
$$\theta_{1}=\theta_{2}=\theta_{3}=\frac{\pi}{2}.$$

The second type of echo occurs when the first two excitation pulses
are identically polarized: $\textbf{l}_{1}$ = $\textbf{l}_{2}$ =
$\textbf{e}_{1}$, while the polarization of the third pulse is
orthogonal to that of the first two: $\textbf{l}_{3}$ =
$\textbf{e}_{2}$. The first two pulses create the non-thermal
velocity distributions of the atoms at the ground state and at the
"bright state" $|b_{1}>$ of the upper level, however the third pulse
couples the ground state to the initially unpopulated orthogonal
"dark" state $|\tilde{b}_{1}>$ of the upper level (\ref{q8}), so
that only the atoms at the ground state contribute to the echo
signal. In this case the echo signal represents itself a "modified"
three-level stimulated photon echo, its polarization coincides with
that of the third excitation pulse:
     \begin{equation}\label{q34}
G_{2}^{e}=\frac{1}{8}\sin(\theta_{1})\sin(\theta_{2})\sin(\theta_{3}),~
G_{1}^{e}=0,
       \end{equation}
while its amplitude constitutes just the half of that of the
stimulated photon echo signal of the first type, in all the other
ways the properties of these first two types of echo signals being
identical.

The third type of echo occurs when the polarizations of the first
two excitation pulses are orthogonal to each other, while the
polarization of the third pulse coincides with that of the first
one: $\textbf{l}_{1}$ = $\textbf{l}_{3}$ = $\textbf{e}_{1}$,
$\textbf{l}_{2}$ = $\textbf{e}_{2}$. In this case the first two
pulses create the coherence on the transition between the "bright"
$|b_{1}>$ and the "dark" $|\tilde{b}_{1}>$ states of the upper
level. The third pulse, which couples the ground state to the
"bright" state $|b_{1}>$, transfers this coherence to the transition
between the ground state and the "dark" state $|\tilde{b}_{1}>$
producing the echo signal on this transition, so that the echo
polarization coincides with that of the second pulse:
       \begin{equation}\label{q35}
G_{2}^{e} =\frac{1}{2}\sin(\theta_{1})
\sin\left(\frac{\theta_{2}}{2}\right)
\sin\left(\frac{\theta_{3}}{2}\right),~ G_{1}^{e}=0.
       \end{equation}
This echo type substantially differs from the first two types, its
maximum amplitude takes place at
$$\theta_{1}=\frac{\pi}{2},~ \theta_{2}=\theta_{3}=\pi,$$
when the signals of the first two types disappear. Generally the
stimulated photon echo contains all the three types of signals, so
that its polarization depends on the areas of the excitation pulses.

Finally, when the polarizations of the first two excitation pulses
are orthogonal to each other and the polarization of the third pulse
coincides with that of the second one: $\textbf{l}_{1}$ =
$\textbf{e}_{1}$, $\textbf{l}_{2}$ = $\textbf{l}_{3}$ =
$\textbf{e}_{2}$, the stimulated echo signal does not appear in the
absence of collisions:
$$F_{n}^{a}=F_{n}^{b}=0.$$
In this case the echo signal will be entirely determined by the
action of the elastic depolarizing collisions, so it may be
identified as collision-induced stimulated photon echo.

Let us now consider the excitation pulses with small areas
$\theta_{n}\ll 1$ and arbitrary polarizations $\textbf{l}_{n}$
($n=1,2,3$). Then we obtain from (\ref{q29})-(\ref{q32}) the
following expression for the echo polarization (unnormalized):
$$
G_{n}^{e}=\frac{1}{8}\theta_{1}\theta_{2}\theta_{3} \left\{
(\textbf{l}_{1}^{*}\textbf{l}_{2})(\textbf{e}_{n}^{*}\textbf{l}_{3})
+
(\textbf{l}_{1}^{*}\textbf{l}_{3})(\textbf{e}_{n}^{*}\textbf{l}_{2})
\right\}.$$ If the second and the third excitation pulses are weak
classical pulses polarized orthogonally to each other:
$\textbf{l}_{2}=\textbf{e}_{1}$, $\textbf{l}_{3}=\textbf{e}_{2}$,
while the first excitation pulse is a single-photon pulse with
arbitrary polarization
$$\textbf{l}_{1}= \xi_{1}\textbf{e}_{1} + \xi_{2}\textbf{e}_{2},~
|\xi_{1}|^{2}+|\xi_{2}|^{2}=1,$$ then the echo pulse is a
single-photon pulse with the polarization
$$\textbf{l}_{e}= \xi_{2}^{*}\textbf{e}_{1} +
\xi_{1}^{*}\textbf{e}_{2}.$$ So, the considered echo scheme may
implement the single q-bit quantum NOT-gate with complex
conjugation, which transforms the q-bit $(\xi_{1},\xi_{2})$, encoded
in the polarization state of the first single-photon excitation
pulse, to the q-bit $(\xi_{2}^{*},\xi_{1}^{*})$, encoded in the
polarization state of the single-photon echo pulse.

\section{Polarization properties of the collision-induced
stimulated photon echo}

The collision-induced stimulated photon echo is formed by the three
excitation pulses, when the polarizations of the first two of them
are orthogonal to each other and the polarization of the third pulse
coincides with that of the second one: $\textbf{l}_{1}$ =
$\textbf{e}_{1}$, $\textbf{l}_{2}$ = $\textbf{l}_{3}$ =
$\textbf{e}_{2}$. Without loss of generality the circular components
of the two unit orthonormal vectors $\textbf{e}_{1}$ and
$\textbf{e}_{2}$ in the plane $XY$ of polarization of the excitation
pulses may be determined by the only one real parameter $\alpha$:
       \begin{equation}\label{q36}
e_{1,q}=\cos(\alpha)\delta_{q,-1}-\sin(\alpha)\delta_{q,1},
       \end{equation}
       \begin{equation}\label{q37}
e_{2,q}=\sin(\alpha)\delta_{q,-1}+\cos(\alpha)\delta_{q,1}.
       \end{equation}
The value $\alpha=0$ corresponds to the circular orthogonal
polarizations -- first pulse right-circularly polarized
$e_{1,q}=\delta_{q,-1}$, second and third -- left-circularly
polarized $e_{2,q}=\delta_{q,1}$, while the value $\alpha=\pi/4$
corresponds to the linear orthogonal polarizations -- first pulse
polarized along the axis $X$, second and third -- along the axis
$Y$. After the substitution of (\ref{q36})-(\ref{q37}) in the
corresponding expressions for the evolution operators the factor
(\ref{q29}) in the echo amplitude becomes as follows:
       \begin{equation}\label{q38}
G_{n}^{e} =\frac{1}{4}\sin(\theta_{1})
\sin\left(\frac{\theta_{2}}{2}\right)
\sin\left(\frac{\theta_{3}}{2}\right) h(\tau_{2}) e^{c}_{n},
       \end{equation}
       \begin{equation}\label{q39}
h(\tau_{2}) = \left(e^{-\Gamma_{b}^{(2)}\tau_{2}}-
e^{-\Gamma_{b}^{(1)}\tau_{2}}\right),
       \end{equation}
       \begin{equation}\label{q40}
e_{1}^{c}=\sin^{2}(2\alpha),~ e_{2}^{c}= -\frac{1}{2}\sin(4\alpha)
\cos\left(\frac{\theta_{3}}{2}\right).
       \end{equation}
In the case of linearly orthogonally polarized excitation pulses
($\alpha = \pi/4$) the echo signal is also linearly polarized like
the first pulse: $e_{1}^{c}= 1$, $e_{2}^{c}=0$, while in the case of
circularly orthogonally polarized excitation pulses ($\alpha = 0$)
the echo signal does not appear: $e_{1}^{c}=e_{2}^{c}=0$. In the
case of circularly polarized pulses the collisional echo does not
appear due to the symmetry with respect to rotation around the
quantization axis. Such a symmetry, as it was shown in \cite{n41},
forbids the collisional conversion of coherence
$\rho^{bb}_{-1,1}(T_{2})$ into $\rho^{bb}_{1,-1}(\tau_{2})$ in the
equation (\ref{q23}) (or $\rho^{bb}_{1,-1}(T_{2})$ into
$\rho^{bb}_{-1,1}(\tau_{2})$), which is needed for the formation of
collisional echo in this case. However, in the case of elliptically
polarized pulses this symmetry is broken and the collisional echo,
which is induced by the collisional transfer of populations
$\rho^{bb}_{-1,-1}(T_{2})$ to $\rho^{bb}_{1,1}(\tau_{2})$ and
$\rho^{bb}_{1,1}(T_{2})$ to $\rho^{bb}_{-1,-1}(\tau_{2})$ in this
case, becomes allowed.

The dependence of the collision-induced echo amplitude on the time
interval $\tau_{2}$ between the second and the third excitation
pulses differs drastically from that of the conventional stimulated
photon echo. The latter decreases exponentially with the increase of
$\tau_{2}$, while the amplitude of the collision-induced echo
increases first from its zero value with the increase of $\tau_{2}$,
then obtains its maximum at
       \begin{equation}\label{q41}
\tau_{2m}=
\frac{\ln(1+\Gamma_{b}^{(2)}/\gamma)-\ln(1+\Gamma_{b}^{(1)}/\gamma)}
{\Gamma_{b}^{(2)}-\Gamma_{b}^{(1)}},
       \end{equation}
and then it exponentially decreases with the further increase of
$\tau_{2}$. The mere existence of the collision-induced stimulated
photon echo is determined by the difference in the alignment
$\Gamma_{b}^{(2)}$ and the orientation $\Gamma_{b}^{(1)}$ relaxation
rates of the excited level. At small $\tau_{2}$ intervals
($|\Gamma_{b}^{(2)}-\Gamma_{b}^{(1)}|\tau_{2}\ll 1$) the echo
amplitude is proportional to $|\Gamma_{b}^{(2)}-\Gamma_{b}^{(1)}|$.

The collision-induced stimulated photon echo, proposed in the
present paper, has much in common with the two-pulse
collision-induced photon echo, observed in ytterbium vapour
\cite{n35}, though there are some essential differences. First, the
two-pulse echo is determined by the difference in the two relaxation
characteristics of the resonant transition, while the stimulated
echo is determined by the difference in the relaxation
characteristics of the excited level. Second, for the existence of
the two-pulse collision-induced echo the dependence of the
relaxation characteristics on the orientation of atomic velocity is
mandatory, while for the existence of the stimulated
collision-induced echo such dependence is insignificant. Third, at
small intervals between the excitation pulses the amplitude of the
two-pulse echo is proportional to the squared difference in two
relaxation rates of the resonant transition, while the amplitude of
the stimulated echo is just proportional to the difference in two
relaxation rates of the excited level, which promises the greater
effect for the stimulated echo, than for the two-pulse echo.

\section{Conclusions}

In the case of simple transition with the angular momentum change
$J_{a}=0 \rightarrow J_{b}=1$ only three atomic states contribute to
the stimulated photon echo formation -- the non-degenerate ground
state and the two substates of the excited level. With the proper
choice of polarizations of excitation pulses three types of echo
signals may be distinguished in the absence of collisions: the
conventional two-level stimulated echo, the modified three-level
stimulated echo and the tri-level echo. Generally the resulting
signal contains all the three types of echoes, so that the echo
polarization depends on the areas of excitation pulses. Such
polarization properties may be employed for the implementation of
quantum NOT-gate. If the polarizations of the second and the third
excitation pulses are identical and orthogonal to that of the first
one, the echo signal is entirely determined by the action of elastic
depolarizing collisions, so it may be identified as
collision-induced stimulated photon echo. The existence of this echo
is determined by the difference in the alignment $\Gamma_{b}^{(2)}$
and the orientation $\Gamma_{b}^{(1)}$ relaxation rates of the
excited level.

In conclusion it should be mentioned, that although the photon echo
formation was treated here under the assumption of narrow spectral
line, when the durations of all excitation pulses were considered to
be much shorter than the inhomogeneous relaxation time, all the
results remain true also in the opposite case of broad spectral
lines, only the dependence on the pulse areas becomes more
complicated than just sine or cosine functions.

{\bf Acknowledgements}

Authors are indebted for financial support of this work to Russian
Foundation for Basic Research (grant 11-02-00141).

\end{document}